\documentclass[reqno,12pt,a4paper]{amsart}

\usepackage{mathrsfs}
\usepackage{amssymb}
\usepackage{amsfonts}
\usepackage{latexsym}
\usepackage{amsthm}

\usepackage{graphicx}
\def\lb{\label}

\newcommand{\er}[1]{\textrm{(\ref{#1})}}

\begin{document}


\renewcommand{\theequation}{\arabic{section}.\arabic{equation}}
\theoremstyle{plain}
\newtheorem{theorem}{\bf Theorem}[section]
\newtheorem{lemma}[theorem]{\bf Lemma}
\newtheorem{corollary}[theorem]{\bf Corollary}
\newtheorem{proposition}[theorem]{\bf Proposition}
\newtheorem{definition}[theorem]{\bf Definition}
\newtheorem{remark}[theorem]{\it Remark}

\def\a{\alpha}  \def\cA{{\mathcal A}}     \def\bA{{\bf A}}  \def\mA{{\mathscr A}}
\def\b{\beta}   \def\cB{{\mathcal B}}     \def\bB{{\bf B}}  \def\mB{{\mathscr B}}
\def\g{\gamma}  \def\cC{{\mathcal C}}     \def\bC{{\bf C}}  \def\mC{{\mathscr C}}
\def\G{\Gamma}  \def\cD{{\mathcal D}}     \def\bD{{\bf D}}  \def\mD{{\mathscr D}}
\def\d{\delta}  \def\cE{{\mathcal E}}     \def\bE{{\bf E}}  \def\mE{{\mathscr E}}
\def\D{\Delta}  \def\cF{{\mathcal F}}     \def\bF{{\bf F}}  \def\mF{{\mathscr F}}
\def\c{\chi}    \def\cG{{\mathcal G}}     \def\bG{{\bf G}}  \def\mG{{\mathscr G}}
\def\z{\zeta}   \def\cH{{\mathcal H}}     \def\bH{{\bf H}}  \def\mH{{\mathscr H}}
\def\e{\eta}    \def\cI{{\mathcal I}}     \def\bI{{\bf I}}  \def\mI{{\mathscr I}}
\def\p{\psi}    \def\cJ{{\mathcal J}}     \def\bJ{{\bf J}}  \def\mJ{{\mathscr J}}
\def\vT{\Theta} \def\cK{{\mathcal K}}     \def\bK{{\bf K}}  \def\mK{{\mathscr K}}
\def\k{\kappa}  \def\cL{{\mathcal L}}     \def\bL{{\bf L}}  \def\mL{{\mathscr L}}
\def\l{\lambda} \def\cM{{\mathcal M}}     \def\bM{{\bf M}}  \def\mM{{\mathscr M}}
\def\L{\Lambda} \def\cN{{\mathcal N}}     \def\bN{{\bf N}}  \def\mN{{\mathscr N}}
\def\m{\mu}     \def\cO{{\mathcal O}}     \def\bO{{\bf O}}  \def\mO{{\mathscr O}}
\def\n{\nu}     \def\cP{{\mathcal P}}     \def\bP{{\bf P}}  \def\mP{{\mathscr P}}
\def\r{\rho}    \def\cQ{{\mathcal Q}}     \def\bQ{{\bf Q}}  \def\mQ{{\mathscr Q}}
\def\s{\sigma}  \def\cR{{\mathcal R}}     \def\bR{{\bf R}}  \def\mR{{\mathscr R}}
\def\S{\Sigma}  \def\cS{{\mathcal S}}     \def\bS{{\bf S}}  \def\mS{{\mathscr S}}
\def\t{\tau}    \def\cT{{\mathcal T}}     \def\bT{{\bf T}}  \def\mT{{\mathscr T}}
\def\f{\phi}    \def\cU{{\mathcal U}}     \def\bU{{\bf U}}  \def\mU{{\mathscr U}}
\def\F{\Phi}    \def\cV{{\mathcal V}}     \def\bV{{\bf V}}  \def\mV{{\mathscr V}}
\def\P{\Psi}    \def\cW{{\mathcal W}}     \def\bW{{\bf W}}  \def\mW{{\mathscr W}}
\def\o{\omega}  \def\cX{{\mathcal X}}     \def\bX{{\bf X}}  \def\mX{{\mathscr X}}
\def\x{\xi}     \def\cY{{\mathcal Y}}     \def\bY{{\bf Y}}  \def\mY{{\mathscr Y}}
\def\X{\Xi}     \def\cZ{{\mathcal Z}}     \def\bZ{{\bf Z}}  \def\mZ{{\mathscr Z}}
\def\O{\Omega}
\def\th{\theta}

\newcommand{\gA}{\mathfrak{A}}
\newcommand{\gB}{\mathfrak{B}}
\newcommand{\gC}{\mathfrak{C}}
\newcommand{\gD}{\mathfrak{D}}
\newcommand{\gE}{\mathfrak{E}}
\newcommand{\gF}{\mathfrak{F}}
\newcommand{\gG}{\mathfrak{G}}
\newcommand{\gH}{\mathfrak{H}}
\newcommand{\gI}{\mathfrak{I}}
\newcommand{\gJ}{\mathfrak{J}}
\newcommand{\gK}{\mathfrak{K}}
\newcommand{\gL}{\mathfrak{L}}
\newcommand{\gM}{\mathfrak{M}}
\newcommand{\gN}{\mathfrak{N}}
\newcommand{\gO}{\mathfrak{O}}
\newcommand{\gP}{\mathfrak{P}}
\newcommand{\gQ}{\mathfrak{Q}}
\newcommand{\gR}{\mathfrak{R}}
\newcommand{\gS}{\mathfrak{S}}
\newcommand{\gT}{\mathfrak{T}}
\newcommand{\gU}{\mathfrak{U}}
\newcommand{\gV}{\mathfrak{V}}
\newcommand{\gW}{\mathfrak{W}}
\newcommand{\gX}{\mathfrak{X}}
\newcommand{\gY}{\mathfrak{Y}}
\newcommand{\gZ}{\mathfrak{Z}}

\def\ve{\varepsilon}   \def\vt{\vartheta}    \def\vp{\varphi}    \def\vk{\varkappa}

\def\Z{{\mathbb Z}}    \def\R{{\mathbb R}}   \def\C{{\mathbb C}}    \def\K{{\mathbb K}}
\def\T{{\mathbb T}}    \def\N{{\mathbb N}}   \def\dD{{\mathbb D}}


\def\la{\leftarrow}              \def\ra{\rightarrow}            \def\Ra{\Rightarrow}
\def\ua{\uparrow}                \def\da{\downarrow}
\def\lra{\leftrightarrow}        \def\Lra{\Leftrightarrow}


\def\lt{\biggl}                  \def\rt{\biggr}
\def\ol{\overline}               \def\wt{\widetilde}
\def\no{\noindent}


\let\ge\geqslant                 \let\le\leqslant
\def\lan{\langle}                \def\ran{\rangle}
\def\/{\over}                    \def\iy{\infty}
\def\sm{\setminus}               \def\es{\emptyset}
\def\ss{\subset}                 \def\ts{\times}
\def\pa{\partial}                \def\os{\oplus}
\def\om{\ominus}                 \def\ev{\equiv}
\def\iint{\int\!\!\!\int}        \def\iintt{\mathop{\int\!\!\int\!\!\dots\!\!\int}\limits}
\def\el2{\ell^{\,2}}             \def\1{1\!\!1}
\def\sh{\sharp}
\def\wh{\widehat}
\def\bs{\backslash}

\def\sh{\mathop{\mathrm{sh}}\nolimits}
\def\all{\mathop{\mathrm{all}}\nolimits}
\def\Area{\mathop{\mathrm{Area}}\nolimits}
\def\arg{\mathop{\mathrm{arg}}\nolimits}
\def\const{\mathop{\mathrm{const}}\nolimits}
\def\det{\mathop{\mathrm{det}}\nolimits}
\def\diag{\mathop{\mathrm{diag}}\nolimits}
\def\diam{\mathop{\mathrm{diam}}\nolimits}
\def\dim{\mathop{\mathrm{dim}}\nolimits}
\def\dist{\mathop{\mathrm{dist}}\nolimits}
\def\Im{\mathop{\mathrm{Im}}\nolimits}
\def\Iso{\mathop{\mathrm{Iso}}\nolimits}
\def\Ker{\mathop{\mathrm{Ker}}\nolimits}
\def\Lip{\mathop{\mathrm{Lip}}\nolimits}
\def\rank{\mathop{\mathrm{rank}}\limits}
\def\Ran{\mathop{\mathrm{Ran}}\nolimits}
\def\Re{\mathop{\mathrm{Re}}\nolimits}
\def\Res{\mathop{\mathrm{Res}}\nolimits}
\def\res{\mathop{\mathrm{res}}\limits}
\def\sign{\mathop{\mathrm{sign}}\nolimits}
\def\span{\mathop{\mathrm{span}}\nolimits}
\def\supp{\mathop{\mathrm{supp}}\nolimits}
\def\Tr{\mathop{\mathrm{Tr}}\nolimits}
\def\BBox{\hspace{1mm}\vrule height6pt width5.5pt depth0pt \hspace{6pt}}
\def\where{\mathop{\mathrm{where}}\nolimits}
\def\as{\text{as}}


\newcommand\nh[2]{\widehat{#1}\vphantom{#1}^{(#2)}}
\def\dia{\diamond}

\def\Oplus{\bigoplus\nolimits}



\def\qqq{\qquad}
\def\qq{\quad}
\let\ge\geqslant
\let\le\leqslant
\let\geq\geqslant
\let\leq\leqslant
\newcommand{\ca}{\begin{cases}}
\newcommand{\ac}{\end{cases}}
\newcommand{\ma}{\begin{pmatrix}}
\newcommand{\am}{\end{pmatrix}}
\renewcommand{\[}{\begin{equation}}
\renewcommand{\]}{\end{equation}}
\def\eq{\begin{equation}}
\def\qe{\end{equation}}
\def\[{\begin{equation}}
\def\bu{\bullet}

\title[{Third order operator with  small periodic coefficients}]
{Third order operator with  small periodic coefficients}

\date{\today}
\author[Andrey Badanin]{Andrey Badanin}
\address{A. Badanin: Northern (Arctic) Federal University,
Russia, e-mail: an.badanin@gmail.com}
\author[Evgeny Korotyaev]{Evgeny Korotyaev}
\address{E. Korotyaev: St.-Petersburg State
University and Pushkin  University, Russia, e-mail:
korotyaev@gmail.com}


\begin{abstract}
We consider the third order operator with   small 1-periodic
coefficients on the real line. The spectrum of the operator is
absolutely continuous and covers all real line. Under the minimal
conditions on the coefficients we show that there are two
possibilities: 1) The spectrum has multiplicity one except for a
small spectral nonempty interval with multiplicity three. Moreover,
the asymptotics of the small interval is determined. 2) All spectrum
has multiplicity one only.

\end{abstract}

\maketitle


\section {Introduction and main results}
\setcounter{equation}{0}

We consider the third order operator $H_\ve$ acting in $L^2(\R)$ and
given by
\[
\lb{Hpq} H_\ve=i\pa^3+\ve V,\qq V=ip\pa+i\pa p+q,
\]
where $\ve\in\R$ is a small coupling constant and the real
1-periodic coefficients $p,q$ satisfy
\[
\lb{pq} 0<\int_0^1 |p(t)|dt+\int_0^1 |q(t)|dt<\iy,\qqq
 \int_0^1 p(t)dt=\int_0^1q(t)dt=0.
\]
Due to \cite{BK3} the operator $H_\ve$ is self-adjoint on the domain
\begin{multline}
\lb{cDH} \mD(H_\ve)=\Bigl\{f\in L^2(\R):i(f''+pf)'+ipf'+qf\in
L^2(\R),\ f'',(f''+pf)'\in L_{loc}^1(\R)\Bigr\},
\end{multline}
the spectrum of $H_\ve$ is absolutely continuous and covers all real
line. Note that the case $p,q\in C^\iy(\R)$ was considered
by McGarvey \cite{McG}. Introduce the
value
\[
\lb{F} h={1\/3}\sum_{n\in\Z}\Bigl({|\wh p_n|^2\/(2\pi n)^2} -{3|\wh
q_n|^2\/(2\pi n)^4}\Bigr),
\]
where $\wh f_n=\int_0^1f(t)e^{-i2\pi nt}dt, n\in\Z$. We present our
main result.

\begin{theorem}
\lb{1.3} i) Let $h>0$. Then there exist two functions $r^\pm(\ve)$,
which are real analytic  in the disk $\{|\ve|<c\}\ss \C$ for some
$c>0$, $r^\pm(0)=0$ and satisfy
\[
\lb{l0} r^+(\ve)-r^-(\ve)=4h^{3\/2}\ve^3+O(\ve^4) \qq\as\qq\ve\to 0.
\]
Moreover, the spectrum of $H_\ve$ has multiplicity one except for a
small spectral nonempty interval $(r^-(\ve),r^+(\ve))$ (or
$(r^+(\ve),r^-(\ve))$) with multiplicity three for any
$\ve\in(-c,c)\sm\{0\}$.

ii) Let $h<0$. Then all spectrum of $H_\ve$ has multiplicity one for
any $\ve\in(-c,c)$.
\end{theorem}

{\bf Remark.} 1) Consider the Hill operator $\cH_\ve=-\pa^2+\ve q$
on the real line. The spectrum of $\cH_\ve$ is absolutely continuous
and consists of spectral bands separated by gaps.
 Recall  that in the case of the
"generic" potential all gaps in the spectrum of $\cH_\ve$ are open,
see \cite{K}  (see also \cite{MO} for the case $q\in L^2(0,1)$).

2) Consider a fourth order operator $\mH_\ve=\pa^4+\ve (\pa p\pa
+q)$ on the real line. Here the real 1-periodic coefficients $p,q$
satisfy \er{pq}  and $\ve\ne 0$ is small enough. The spectrum of
$\mH_\ve$ is absolutely continuous and consists of spectral bands
separated by gaps.  The authors proved that if either $p=0,q\ne 0$
\cite{BK1} or $q=0,p\ne 0$ \cite{BK2}, then there exists a small
non-empty spectral interval with the spectrum of multiplicity 4 and
all other spectral spectrum have  multiplicity 2.

Note that the result for the operator $\mH_\ve$ in the general case
$p\ne 0,q\ne 0, \ve \to 0$ is more complicated. One can choose $p,q$
such that all the spectrum has multiplicity two. It follows from the
Papanicolaou result \cite{P} that any point of the spectrum of the
Euler-Bernoulli equation $(\x y'')''=\l \e y$ with any periodic
$\x>0,\e>0$, has multiplicity 2.

3) If $q=0$, then for real  $\ve\ne0$ small enough there exists a
small non-empty spectral band in $\s(H_\ve)$ with the spectrum of
multiplicity 3. It is similar to the case of the operator $\mH_\ve$.

4) If $p=0$, then for $\ve\ne0$ small enough all spectrum of $H_\ve$
has multiplicity one. This is in contrast with the case of the
fourth order operator $\mH_\ve$.

5) For the operator $H_\ve$ the endpoints of the spectral interval
with multiplicity 3 are branch points of the multipliers, see
\er{sro}. For the operator $\mH_\ve$  one endpoint of the spectral
interval with multiplicity 4 is a branch point and another endpoint
is a periodic eigenvalue.

6) The proof of the theorem is based on the analysis of the
monodromy matrix as $\ve\to 0$ and  identities \er{rtr}, \er{sro}.
In order to show \er{l0}  we determine the asymptotics of
$r^\pm(\ve)$ in the form $r^\pm(\ve)=r(\ve)\pm
2h^{3\/2}\ve^3+O(\ve^4)$ as $\ve\to 0$, where $r$ is some function.
This gives the asymptotics of $r^+(\ve)-r^-(\ve)$.

The operator $i\pa^3+ip\pa+i\pa p+q$ is used in the inverse problem
method of integration of the non-linear evolution Boussinesq
equation, see, e.g., \cite{McK}.  The scattering and inverse
scattering theory for operators $i\pa^3+ip\pa+i\pa p+q$ with
decreasing potentials was developed in \cite{DTT}.  The work of
McKean \cite{McK} give the numerous results in the inverse spectral
theory for the non-self-adjoint operator $\pa^3+p\pa+\pa p+q$ (and
also for the multi-point Dirichlet problem), where $p$ and $q$ are
smooth and sufficiently small. The third order operator on the
bounded interval was considered in \cite{A1}, \cite{A2}. The
spectral properties of the arbitrary order operator with smooth
periodic coefficients are studied in \cite{DS}. Note that the fourth
order operator $\pa^4+\pa p\pa +q$ is also used for the integration
of some non-linear evolution equation, see \cite{HLO}.

In \cite{BK3} we study the basic spectral properties of the third
order operator  $H_\ve$ at $\ve =1$. We introduce the Lyapunov
function $\D$, which is analytic on a 3-sheeted Riemann surface
$\mR$ and satisfies the usual identity $\D=\cos k$, where $k$ is a
quasimomentum. This important property essentially complicates the
spectral analysis of the operator $H_\ve$. Recall that the Lyapunov
function for the second order operator $\cH_\ve$ is entire and the
Riemann surface for the fourth order operator $\mH_\ve$ is
2-sheeted, see \cite{BK1}, \cite{BK2}. Using the
direct integral decomposition we prove that the spectrum of $H_\ve$
is absolutely continuous and covers the whole real line. Moreover,
we describe the spectrum, counting with multiplicity, in terms of
the Lyapunov function similarly to the case of the Hill operator. In
\cite{BK4} we determine the high energy asymptotics of the periodic
and antiperiodic spectrum and the branch points of the Lyapunov
function.

\section {Monodromy matrix}
\setcounter{equation}{0}

We consider the equation
\[
\lb{1b} i y'''+ \ve Vy=\l y,\qqq \l\in\C.
\]
We introduce the $3\ts
3$ matrix-valued function $M(t,\l,\ve), (t,\l,\ve)\in\R\ts\C\ts\R,$ by
\[
\lb{deM}
M(t,\l,\ve)=\ma\vp_1&\vp_2&\vp_3\\
\vp_1'&\vp_2'&\vp_3'\\
\vp_1''+\ve p\vp_1&\vp_2''+\ve p\vp_2&\vp_3''+\ve p\vp_3\am(t,\l,\ve),\qq
\]
where $\vp_j,j=1,2,3$, are the fundamental solutions of equation \er{1b}
satisfying the conditions
\[
\lb{ic}
M(0,\l,\ve)=\1_3,
\]
and $\1_3$ is the $3\ts 3$ identity matrix.
Due to \er{1b}, \er{deM} the matrix-valued function
$M(t,\l,\ve)$ satisfies the matrix equation
\[
\lb{me1} M'-P(\l)M=\ve Q(t)M,\qqq
(t,\l,\ve)\in\R\ts\C\ts\R,
\]
where the $3\ts 3$ matrices $P$ and $Q$ are given by
\[
\lb{mu}
P=\ma 0&1&0\\0 &0&1\\
-i\l&0&0\am,\qqq Q=\ma 0&0&0\\-p&0&0\\iq&-p&0\am.
\]
Due to $Q\in L^1(\T)$  there exists absolutely continuous in
$t\in\R$ matrix-valued solution $M(t,\l,\ve)$  of the problem
\er{me1}--\er{ic} and {\it the monodromy matrix}
$M(1,\l,\ve),(\l,\ve)\in\C\ts\R$, is well-defined.

Consider the case $\ve =0$.
The matrix-valued solution $M_0(t,\l)=M(t,\l,0)$ of the problem
\er{me1}--\er{ic}
has the form
$
M_0
=e^{tP}.
$
Each function $M_0(t,\cdot), t\in\R$, is entire.
Eigenvalues of the matrix $P$ are given by $iz,i\o z,i\o^2 z$,
henceforth
$$
\o=e^{i{2\pi\/3}},\qq z=\l^{1\/3},\qq
\arg\l\in\Bigl(-{\pi\/2},{3\pi\/2}\Bigr], \qq\arg
z\in\Bigl(-{\pi\/6},{\pi\/2}\Bigr].
$$
Then eigenvalues of the matrix $M_0$ have the form $e^{izt},e^{i\o zt},e^{i\o^2 zt}$.
Estimates
$|e^{iz\o^jt}|\le e^{z_0|t|}$ imply
\[
\lb{eu1}
|M_0(t,\l)|\le e^{z_0|t|},\qqq\text{all}\qq
(t,\l)\in\R\ts\C,\qq\where\qq
z_0=\max_{j=0,1,2}\Re(iz\o^j)=\Re(iz\o^2),
\]
and henceforth
a matrix $A$ has the norm given by
$$
|A|=\max\{\sqrt{f}:f\ \text{is an eigenvalue of the matrix}\ A^*A\}.
$$

Consider the case $\ve \ne 0$.
Using the standard arguments we deduce that the function $M(t,\l,\ve)$
satisfies the integral equation
\[
\lb{iev}
M(t,\l,\ve)=M_0(t,\l)+\ve\int_0^tM_0(t-s,\l)Q(s)M(s,\l,\ve)ds.
\]
The standard iterations in \er{iev} lead to the formal series
\[
\lb{evj}
M(t,\l,\ve)=\sum_{n\ge 0}\ve^nM_n(t,\l),\qq
M_n(t,\l)=\int_0^tM_0(t-s,\l)Q(s)M_{n-1}(s,\l)ds,\qq n\ge 1.
\]

In the following Lemma we will prove estimates of the monodromy matrix $M(1,\l,\ve)$.

\begin{lemma}
\lb{T21} Each matrix-valued function $M(t,\cdot,\cdot), t\in [0,1]$ is
entire in $(\l,\ve)\in \C^2$ and satisfies:
\[
\lb{2if}
|M(1,\l,\ve)|\le e^{z_0+\vk},\qq
|M(1,\l,\ve)-\sum_{n=0}^{N-1}M_n(1,\l,\ve)|\le |\ve|^N\vk^Ne^{z_0+|\ve|\vk},\qq
\]
for all $(N,\l,\ve)\in\N\ts\C^2$, where
$$
\vk=\int_0^1\bigl(|p(t)|+|q(t)|\bigr)dt.
$$

\end{lemma}

\no {\bf Proof.}  Identity \er{evj} gives
\[
\lb{2ig} M_n(t,\l)=\int\limits_{0< t_1<...<
t_n<t_{n+1}=t}\prod\limits_{k=1}^{n}
\rt(M_0(t_{k+1}-t_k,\l)Q(t_k)\rt)M_0(t_1,\l)dt_1dt_2...dt_n,
\]
the factors are ordering from right to left.
Substituting estimates \er{eu1} into identities
\er{2ig} we obtain
\[
\lb{eM1} |M_n(t,\l)|\le{e^{z_0t}\/n!}\lt(\int_0^t|Q(s)|ds\rt)^n,
\qqq \all \qq (n,t,\l)\in\N\ts\R_+\ts\C.
\]
These estimates show that for each fixed $t\ge0 $ the formal series
\er{evj} converges absolutely and uniformly on bounded subset of
$\C^2$. Each term of this series is an entire function of
$(\l,\ve)$. Hence the sum is an entire function. Summing the
majorants we get
$$
|M(t,\l,\ve)-\sum_{n=0}^{N-1}\ve^nM_n(t,\l)|\le
\lt(|\ve|\int_0^t|Q(s)|ds\rt)^N e^{z_0t+|\ve|\int_0^t|Q(s)|ds}
$$
for all $(t,\l,\ve)\in\R_+\ts\C^2$, which yields \er{2if}.
$\BBox$

The characteristic polynomial $D$ of the monodromy matrix $M(1,\l,\ve)$ is given by
\[
\lb{1c} D(\t,\l,\ve)=\det(M(1,\l,\ve)-\t \1_{3}),\qq (\t,\l,\ve)\in\C^3.
\]
An eigenvalue
of $M(1,\l,\ve)$ is called a {\it multiplier}, it is a zero of the
algebraic equation $D(\cdot,\l,\ve)=0$. Each $M(1,\l,\ve), (\l,\ve)\in\C\ts\R$,
has exactly $3$ (counted with multiplicities)  multipliers
$\t_j(\l,\ve),j=1,2,3$. We need following results.

\begin{lemma} \lb{TMM}
i) The function $D$ satisfies
\[
\lb{cM} D(\t,\l,\ve)=-\t^3+\t^2T(\l,\ve)-\t\ol
T(\ol\l,\ve)+1,\qq\all\qq(\t,\l,\ve)\in\C^2\ts\R,
\]
where
\[
\lb{defT} T(\l,\ve)=\Tr M(1,\l,\ve).
\]
If $\t(\l,\ve),(\l,\ve)\in\C\ts\R$, is a multiplier, then $\bar\t^{-1}(\bar\l,\ve)$ is also a
multiplier.

ii) Let $(\l,\ve)\in\R^2$. Two cases are possible only:

\no a) all three multipliers belong to the unit circle;

\no b) exactly one simple multiplier belongs to the unit circle.

\no In the case b) the multipliers have the form
\[
\lb{mrs} e^{i k},\qq e^{i\bar k}, \qq e^{-i2\Re k},\qq\text{some}
\qq   k\in\C:\Im k\ne 0.
\]

iii) The spectrum $\s(H_\ve)$ of the operator $H_\ve,\ve\in\R$, is absolutely
continuous and satisfies
\[
\lb{sdD} \s(H_\ve)=\{\l\in\R:|\t_j(\l,\ve)|=1,\ \text{some}\ j=1,2,3\}.
\]
The spectrum has multiplicity 3 (in the case ii a)  or 1 (in the case ii b).

\end{lemma}

\no {\bf Proof.} Lemma have been proved in \cite{BK3}, we give the
sketch of the proof for completeness.

i) Identity \er{me1} yields $JM'=V M$ and $-( M^*)'J= M^*V$ for
$\l\in\R$, where
$$
J=\ma
0&0&i\\
0&-i&0\\
i&0&0 \am,\qq
V=J(P+Q)
=\ma \l-q&-ip&0\\ip&0&-i\\0&i&0\am.
$$
Then $ ( M^*J M)'=- M^*V M+ M^*V M=0, $ which yields
\[
\lb{idM}
M^*(1,\ol\l)J M(1,\l)=J,
\]
here and below in this proof $M(t,\l)=M(t,\l,\ve),...$
Identity \er{deM} and equation \er{1b} give
$
(\det M)'=0.
$
Then $\det M(t,\l)=\det M(0,\l)=1$, for all
$(t,\l)\in\R\ts\C$, which yields
$ \det M(1,\l)=1$.
Direct calculations show that
$$
D(\t,\l)=\det ( M(1,\l)-\t \1_3)=-\t^3+\t^2\Tr M(1,\l)+B(\l)\t-1
\qq\all\qq(\t,\l)\in\C^2,
$$
where $B(\l)=\pa_\t D(0,\l)$. The standard formula from the matrix
theory gives
$$
\pa_\t D(\t,\l)=-D(\t,\l)\Tr( M(1,\l)-\t\1_3)^{-1}.
$$
Using the identity $D(0,\l)=1$, we obtain $B(\l)=-\Tr
M^{-1}(1,\l)$. Identity \er{idM} gives $ M^{-1}(1,\l)=-J
M^*(1,\ol\l)J$, which implies $\Tr M^{-1}(1,\l)=\Tr M^*(1,\ol\l)$,
for all $\l\in\R$. Then $B(\l)=-\Tr M^*(1,\ol\l)$, which yields
\er{cM}.

ii) Identity \er{cM} implies
\[
\lb{dM}
D(\t,\l)=-\t^3 { \ol D(\bar\t^{-1},\l)}\qq\text{дл€ всех}\qq
(\t,\l)\in\C\ts\R,\qq \t\ne 0.
\]
Then if $\t(\l)$ is a zero of $D(\t,\l)$ for some
$\l\in\R$, then $\bar\t^{-1}(\l)$ is also a zero.
Using the identity $\t_1\t_2\t_3=\det M(1,\cdot)=1$, we obtain
the needed statements.

iii) The proof is standard and uses the direct integral
decomposition of the operator $H_\ve$, see \cite{RS}. $\BBox$

Fix $\ve\in\R$. The coefficients of the polynomial $D(\cdot,\l,\ve)$
are entire functions in $\l$. It is known (see, e.g., \cite{Fo},
Ch.~8) that the roots $\t_j(\cdot,\ve),j=1,2,3$, constitute one, two
or three branches of one, two or three analytic functions that have
only algebraic singularities in $\C$. The simple analysis (see
\cite{McK}) shows that the functions $\t_j(\cdot,\ve)$ constitute
three branches of one function $\t(\cdot,\ve)$ analytic on a
3-sheeted Riemann surface. Asymptotics \er{ars}, proved in the
following Lemma, shows that the functions $\t_j(\cdot,\ve)$ are all
distinct, that is if $k\ne \ell$, then $\t_j(\l,\ve)\ne
\t_\ell(\l,\ve)$ for all $\l\in\C$ with the exception of some
special values of $\l$. There are only a finite number of such {\it
exceptional points} in any bounded domain.

We introduce the {\it discriminant} $\r(\l,\ve),(\l,\ve)\in\C^2$, of the polynomial
$D(\cdot,\l,\ve)$ by
\[
\lb{r} \r=(\t_1-\t_2)^2(\t_1-\t_3)^2(\t_2-\t_3)^2.
\]

Consider the case $\ve=0$. The multipliers $\t_j^0$, the trace $T_0$
of the monodromy matrix and the discriminant $\r_0(\l)=\r(\l, 0)$
are given by
\[
\begin{aligned}
\lb{t01}
\t_1^0=e^{iz},\qq \t_2^0=e^{i\o z},\qq \t_3^0=e^{i\o^2 z},
 \qqq \o=e^{i{2\pi\/3}},\\
 T_0(\l)=\Tr M_0(1,\l)=e^{iz}+e^{i\o z}+e^{i\o^2 z},
\end{aligned}
\]
and
\[
\lb{ro0} \r_0=(e^{iz}-e^{i\o z})^2(e^{iz}-e^{i\o^2 z})^2(e^{i\o z}-e^{i\o^2 z})^2
=64\sinh^2{\sqrt 3 z\/2}\sinh^2{\sqrt 3 \o z\/2}
\sinh^2{\sqrt 3 \o^2z\/2}.
\]

\begin{lemma}\lb{Thr}
i) The function $\r(\l,\ve)$ is entire, real on $(\l,\ve)\in\R^2$, and  satisfies:
\[
\lb{rtr} \r(\l,\ve)= |T(\l,\ve)|^4-8\Re T^3(\l,\ve)+18|T(\l,\ve)|^2-27, \qq
\text{all}\qq(\l,\ve)\in\R^2,
\]
\[
\lb{ars} \r(\l,\ve)=\r_0(\l)\bigl(1+O(\ve)\bigr) \qq\as\qq\ve\to 0.
\]
uniformly in $\l$ on any bounded subset of
$\cD=\C\sm\cup_{n\in\Z}\{|\l-i({2\pi n\/\sqrt3})^3|\le 1\}$.

ii) Let $\ve\in\R$ and let $\gS_3$ be the part of the spectrum of $H_\ve$ having the
multiplicity 3. Then
\[
\lb{sro} \gS_3=\{\l\in\R:|\t_j(\l,\ve)|=1,\ \text{for all}\
j=1,2,3\}=\{\l\in\R:\r(\l,\ve)\le 0\}.
\]

iii)  Let  $|\ve|<c$ for some $c>0$ small enough. Then the function
$\r(\cdot,\ve )$ has exactly two zeros, counted with multiplicities,
in each domain $|\l-i({2\pi n\/\sqrt3})^3|<1,n\in\Z$. There are no
other zeros.
In particular, the function $\r(\cdot,\ve)$ has no any real zeros in
the domain $|\l|\ge 1$.

\end{lemma}

\no {\bf Proof.}
i) The function $\r$ is a discriminant of the cubic polynomial \er{cM}
with entire coefficients, then $\r$ is
entire.
The standard formula for the discriminant of the polynomial
$-\t^3+a\t^2-b\t+1$ gives $d=(ab)^2-4(a^3+b^3)+18ab-27$,
which yields \er{rtr}.

We will show \er{ars}. Estimates \er{2if} give
$M(1,\l,\ve)=M_0(1,\l)+O(\ve)$ as $\ve\to 0$, uniformly in $\l$ on
any compact in $\C$. Let $\l\in\cD$. Then  all eigenvalues
$e^{iz},e^{i\o z},e^{i\o^2 z}$ of the matrix $M_0(1,\l)$ are simple
and the standard matrix perturbation theory gives
$$
\t_j(\l,\ve)=\t_j^0(\l)\bigl(1+O(\ve)\bigr),\qq\as\qq \ve\to 0,
$$
uniformly in $\l$ on any bounded subset of
$\cD$.
Substituting these asymptotics into \er{r} we obtain
\er{ars}.

ii) Lemma \ref{TMM} iii) implies the first identity in \er{sro}.
Let $\ve\in\R$ and let $k_j,j=1,2,3$, be given by $\t_j=e^{ik_j}$.
If $k\ne \ell$, then $k_j(\l,\ve)\ne k_\ell(\l,\ve)$ for all
non-exceptional $\l\in\C$. Identity $\t_1\t_2\t_3=1$ yields
$k_1+k_2+k_3=0$. This
identity and \er{r} give
\[
\lb{rqm}
\r=(e^{ik_1}-e^{ik_2})^2
(e^{ik_1}-e^{ik_3})^2
(e^{ik_2}-e^{ik_3})^2
=-64\sin^2{k_1-k_2\/2}
\sin^2{k_1-k_3\/2}
\sin^2{k_2-k_3\/2}.
\]
If $\l\in\gS_3$, then, due to the first identity in \er{sro}, $k_j(\l,\ve)\in\R$
for all $j=1,2,3$, and \er{rqm} yields $\r(\l,\ve)\le 0$.
If $\l\in\s(H_\ve)\sm\gS_3$, then exactly one $k_j$, say $k_1$,
is real and $k_2=\ol k_3$ are non-real, see \er{mrs}. Identity \er{rqm} implies $\r(\l,\ve)>0$,
which yields the second identity in \er{sro}.

iii) Let $\l$ belong to the
contours $|\l-i({2\pi n\/\sqrt3})^3|=1,n\in\Z$.
Asymptotics \er{ars} yields
$$
|\r(\l,\ve)-\r_0(\l)|=|\r_0(\l)|\Bigl|{\r(\l,\ve)\/\r_0(\l)}-1\Bigr|
=|\r_0(\l)|O(\ve)<|\r_0(\l)|
$$
on all contours.
Hence, by Rouch\'e's theorem, $\r$ has as many zeros,
as $\r_0$ in each of the
bounded domains and the remaining unbounded domain. Since
$\r_0$ has exactly one zero of multiplicity two
at each $i({2\pi n\/\sqrt 3})^3,n\in\Z$,
the statement follows.
$\BBox$

\section{Proof of the main Theorem }
\setcounter{equation}{0}

Introduce the entire functions
$T_n=\Tr M_n(1,\cdot),n\ge 0.$
Estimates \er{2if} imply
\[
\lb{e2}
T(\l,\ve)=T_0(\l)+\ve T_1(\l)+\ve^2 T_2(\l)+\ve^3 T_3(\l)+O(\ve^4)
\qq\as\qq\ve\to 0
\]
uniformly in $\l$ on any compact in $\C$.

\begin{lemma}
Let  $h$ be given by \er{F}. Then the  following relations hold
true:
\[
\lb{T0}
T_1=0,
\]
\[
\lb{T1} \Re T_2(\l)=-3h(1+O(\l))\qq\as\qq\l\to 0.
\]
\end{lemma}

\no {\bf Proof.}  Identity \er{evj} and conditions \er{pq} imply
$$
T_1=\Tr M_1(1,\cdot)=\Tr\int_0^1 e^{(1-t)P}Q(t)e^{tP}ds =\Tr e^{P}
\int_0^1 Q(t)dt=0,
$$
which yields \er{T0}.
Moreover,
\[
\lb{T2} T_2=\Tr M_2(1,\cdot) =\Tr\int_0^1\int_0^t
e^{(1-t+s)P}Q(t)e^{(t-s)P}Q(s)dsdt.
\]
We rewrite identities \er{me1} for
the matrices $P,Q$ in the form
$$
P=P_0-i\l P_1,\qq Q=-pP_0^*+i qP_1,\qq P_0=\ma 0&1&0\\0 &0&1\\
0&0&0\am,\qq
P_1=\ma 0&0&0\\0 &0&0\\ 1&0&0\am.
$$
Using
the identities
$$
e^{tP_0}=\1_3+tP_0+{t^2\/2}P_0^2=
\ma1& t& {t^2\/2}\\
0&1&t\\
0&0&1\am,
$$
we obtain
$$
e^{tP(\l)}Q(s)=e^{tP_0}Q(s)(\1_3+O(\l))
=(-p(s)K_1(t)+iq(s)K_2(t))(\1_3+O(\l))
$$
as $\l\to 0$, uniformly on $(t,s)\in[0,1]^2$, where
$$
K_1=e^{tP_0}P_0^*=\ma t& {t^2\/2}& 0\\
1&t&0\\
0&1&0\am,\qq K_2=e^{tP_0}P_1=\ma {t^2\/2}&0& 0\\
t&0&0\\
1&0&0\am.
$$
Using the identities
$$
\Tr K_1(1-u) K_1(u)=2(1-u)u+{(1-u)^2\/2}+{u^2\/2} ={1\/2}+u(1-u),
$$
$$
\Tr K_2(1-u) K_2(u)={(1-u)^2u^2\/4},
$$
we obtain
$$
\Re\Tr e^{(1-t+s)P}Q(t)e^{(t-s)P}Q(s)=\F(t,s)(1+O(\l)),
$$
as $\l\to 0$, uniformly on $(t,s)\in[0,1]^2$, where
$$
\F(t,s)=p(t)p(s)\Bigl({1\/2}+u(1-u)\Bigr) -q(t)q(s){(1-u)^2u^2\/4},
\qq u=t-s.
$$
Substituting this asymptotics into \er{T2} we obtain
\[
\lb{idA}
\Re T_2(\l) =\int_0^1dt\int_0^t \F(t,s)ds(1+O(\l))
={1\/2}\int_0^1dt\int_{t-1}^t \F(t,s) ds(1+O(\l))
\]
as $\l\to 0$, since $\F(t,s)=\F(s,t-1)$. Using  the simple identity
$$
\int_0^1dt\int_{t-1}^tg(t-s)f(t)f(s)ds =\sum_{k\in\Z}|\wh
f_k|^2\wh g_k,
$$
for all $f\in L^1(0,1), g,g'\in L^2(0,1), g(0)=g(1)=0$, and
identities
$$
\int_0^1u(1-u)e^{-i2\pi nu}du=-{1\/2(\pi n)^2},\qqq
\int_0^1u^2(1-u)^2e^{-i2\pi nu}du=-{3\/2(\pi n)^4},
$$
for all $n\ne 0$, identities \er{idA} imply \er{T1}. $\BBox$

\no {\bf Proof of Theorem \ref{1.3}.} i) Identities \er{t01} imply
\[
\lb{e1}
T_0(\l)=e^{iz}+e^{i\o z}+e^{i\o^2z}=3-{i\l\/2}-{\l^2\/240}+O(\l^3)\qq\as\qq\l\to 0.
\]
Recall that the functions $T(\l,\ve),\r(\l,\ve)$ are entire in $(\l,\ve)\in\C^2$.
Let the entire functions $a(\l,\ve),b(\l,\ve)$
and the numbers $b_j,j=2,3$, be given by
$$
T(\l,\ve)=3+a(\l,\ve)+ib(\l,\ve),\qq b(\l,\ve)=\Im T(\l,\ve),\qq\all\qq
(\l,\ve)\in\R^2,\qq b_j=\Im T_j(0).
$$
Substituting relations \er{T0}, \er{T1}, \er{e1} into \er{e2} we obtain
\[
\lb{sp1} T(\l,\ve)=3-{i\l\/2}-{\l^2\/240}
-\bigl(3h-ib_2\bigr)\ve^2+\ve^3T_3(0)+O(\l^3)+O(\l\ve^2)+O(\ve^4)
\]
as $(\l,\ve)\to(0,0)$.
Asymptotics \er{sp1} gives
\[
\lb{sp5}
a(\l,\ve)=-3h\ve^2+O(\l^2)+O(\l\ve^2)+O(\ve^3),\qq
b(\l,\ve)=\m(\l,\ve)+O(\l^3)+O(\l\ve^2)+O(\ve^4),
\]
as $(\l,\ve)\to (0,0)$, where
\[
\lb{e4}
\m=-{\l\/2}+{r\/2},\qqq r=2b_2\ve^2+2b_3\ve^3.
\]
Identity \er{rtr} gives
\[
\lb{sp4} \r=a^3(a+4)+b^2\bigl(108+2(a+18)a+b^2\bigr).
\]
Substituting asymptotics \er{sp5} into \er{sp4} we obtain
$$
\r(\l,\ve)=f(\m,\ve)
=108\Bigl(\m^2-h^3\ve^6+O(\m^4)+O(\m^2\ve^2)+O(\m\ve^4)+O(\ve^7)\Bigr),
$$
as $(\m,\ve)\to(0,0)$,  where
\[
\lb{sc2}
\l=r-2\m,
\]
and $f(\m,\ve)$ is entire in $(\m,\ve)$. Introduce the
new variable $u$ by $\m=u\ve^3$. Then
$$
f(u\ve^3,\ve)=108\ve^6 E(u,\ve),
$$
where $E(u,\ve)$ is entire in $(u,\ve)$ and
$$
E(u,\ve) =u^2-h^3
+O(\ve),
$$
as $\ve\to 0$ uniformly in $u$.

Consider the equation $E(u,\ve)=0$. Let $h\ne 0$. Using ${\pa\/\pa u}E(\pm
h^{3\/2},0)\ne 0$ and the Implicit Function Theorem we deduce that
there exist two real analytic functions $u^\pm(\ve)$ in the disk
$\{|\ve|<c\}$ for some $c>0$ such that each $u^{\pm}(\ve)$ is a
zero of the function $E(\cdot,\ve)$. These functions satisfy
$$
u^\pm(\ve)=\pm h^{3\/2}+O(\ve)\qq\as\qq\ve\to 0.
$$
Substituting $\mu=u^\pm(\ve)\ve^3$ into the identity \er{sc2}
we deduce that there exist two
real analytic functions $r^\pm(\ve)$ in the disk $\{|\ve|<c\}$ such
that
$$
r^\pm(\ve)=r(\ve)\pm 2h^{3\/2}\ve^3+O(\ve^4),\qq\as\qq\ve\to 0,
$$
which yields \er{l0}.

Consider $h>0$. Let $\ve>0$ be small enough. Then
$r^-(\ve)<r(\ve)<r^+(\ve)$. Asymptotics \er{sc2}
shows that $\r(r(\ve),\ve)=f(0,\ve)<0$. Then $\r(\cdot,\ve)<0$
on the whole interval $(r^-(\ve),r^+(\ve))$ and, by Lemma \ref{Thr}
iii), $\r(\cdot,\ve)\ge 0$ out of this interval. Then, due to
\er{sro}, the spectrum of $H_\ve$ in this interval has multiplicity
3 and the other spectrum has multiplicity 1. The proof for the case
$\ve<0$ is similar.

ii) If $h<0$, then, by Lemma \ref{Thr} iii), the function $\r$
has no any real zeros and $\r(\cdot,\ve)>0$ on the whole real axis.
Hence all the spectrum has multiplicity 1. $\BBox$

\no {\bf Acknowledgments.} \small This work was supported by the
Ministry of education and science of the Russian Federation, state
contract 14.740.11.0581.


\begin{thebibliography}{9999}
\setlength{\itemsep}{-\parskip} \footnotesize

\bibitem[A1]{A1} Amour, L. Determination of a third-order operator from two of its spectra,
SIAM J. Math. Anal., 30 (1999), 1010Ц-1028.

\bibitem[A2]{A2} Amour, L. Isospectral flows of third order operators.
SIAM J. Math. Anal., 32 (2001), no. 6, 1375Ц-1389.

\bibitem[BK1]{BK1} Badanin, A.; Korotyaev, E.
Spectral asymptotics  for periodic fourth-order operators, Int.
Math. Res. Not. , 45 (2005), 2775--2814.

\bibitem[BK2]{BK2} Badanin, A.; Korotyaev, E.
Spectral estimates for periodic fourth order operators. Algebra i
Analiz, 22:5 (2010), 1-48.


\bibitem[BK3]{BK3} Badanin, A.; Korotyaev, E.
Third order operator  with periodic coefficients on the real line,
 preprint, 2011.

\bibitem[BK4]{BK4} Badanin, A.; Korotyaev, E.
Spectral asymptotics  for third order operator with periodic
coefficients, preprint,  2011.



\bibitem[DTT]{DTT} Deift, P.; Tomei, C.; Trubowitz, E.
Inverse scattering and the Boussinesq equation, Comm. on Pure and
Appl. Math., 35 (1982), 567--628.

\bibitem[DS]{DS}
Dunford, N.; Schwartz, J. Linear Operators, Part II: Spectral
Theory, Interscience, New York, 1988.

\bibitem[Fo]{Fo} Forster, O. Lectures on Riemann surfaces.
Graduate Texts in Mathematics, 81, Springer -Verlag, New York, 1991.

\bibitem[HLO]{HLO} Hoppe, J.; Laptev, A.; \"Ostensson, J.
Solitons and the removal of eigenvalues for fourth-order
differential operators,  Int. Math. Res. Not., 2006, 14 pp.


\bibitem[K]{K}
Korotyaev,  E. Characterization of the spectrum of Schr\"odinger operators with
periodic distributions. Int. Math. Res. Not., 37 (2003), 2019-2031.

\bibitem[MO]{MO} Marchenko, V. A., Ostrovskii I. V.
Characteristics of the spectrum of the Hill operator. (Russian) Mat.
USSR Sb., 26 (1975), no.4, 493-554.

\bibitem[McG]{McG} McGarvey, D. C.
Differential operators with periodic coefficients in $L_p(-\iy,\iy)$,
J. Math. Anal. Appl. 11 (1965), 564Ц-596.

\bibitem[McK]{McK} McKean, H. Boussinesq's equation on the circle, Com. Pure and
Appl. Math., 34(1981), 599--691.


\bibitem[P]{P}  Papanicolaou V.
The Spectral Theory of the Vibrating Periodic Beam, Commun. Math.
Phys. 170(1995), 359 -- 373.

\bibitem[RS]{RS} Reed, M.;  Simon, B. Methods of modern mathematical
physics. IV. Analysis of operators, Academic Press, New York-London,
1978.




\end{thebibliography}
\end{document}